\documentstyle[11pt,paspconf,epsf]{article}

\def\plotone#1{\centering \leavevmode
\epsfxsize=\columnwidth \epsfbox{#1}}

\begin{document}

\title{The Future of Radio Astronomy: Options for Dealing
with Human Generated Interference}

\author{R. D. Ekers and J. F. Bell}
\affil{ATNF CSIRO, PO Box 76 Epping NSW 1710, Sydney Australia; rekers@atnf.csiro.au~~~jbell@atnf.csiro.au}




\begin{abstract}
Radio astronomy provides a unique window on the universe, allowing us
to study: non-thermal processes (galactic nuclei, quasars, pulsars) at
the highest angular resolution using VLBI, with low opacity. It is the
most interesting wave band for SETI searches. To date it has yielded 3
Nobel prizes (microwave background, pulsars, gravitational radiation).
There are both exciting possibilities and substantial challenges for
radio astronomy to remain at the cutting edge over the next 3
decades. New instruments like ALMA and the SKA will open up new
science if the challenge of dealing human generated interference can
be met. We summarise some of the issues and technological developments
that will be essential to the future success of radio astronomy.
\end{abstract}




\section{Telescope Sensitivity}

\begin{figure}[htbp]
\plotone{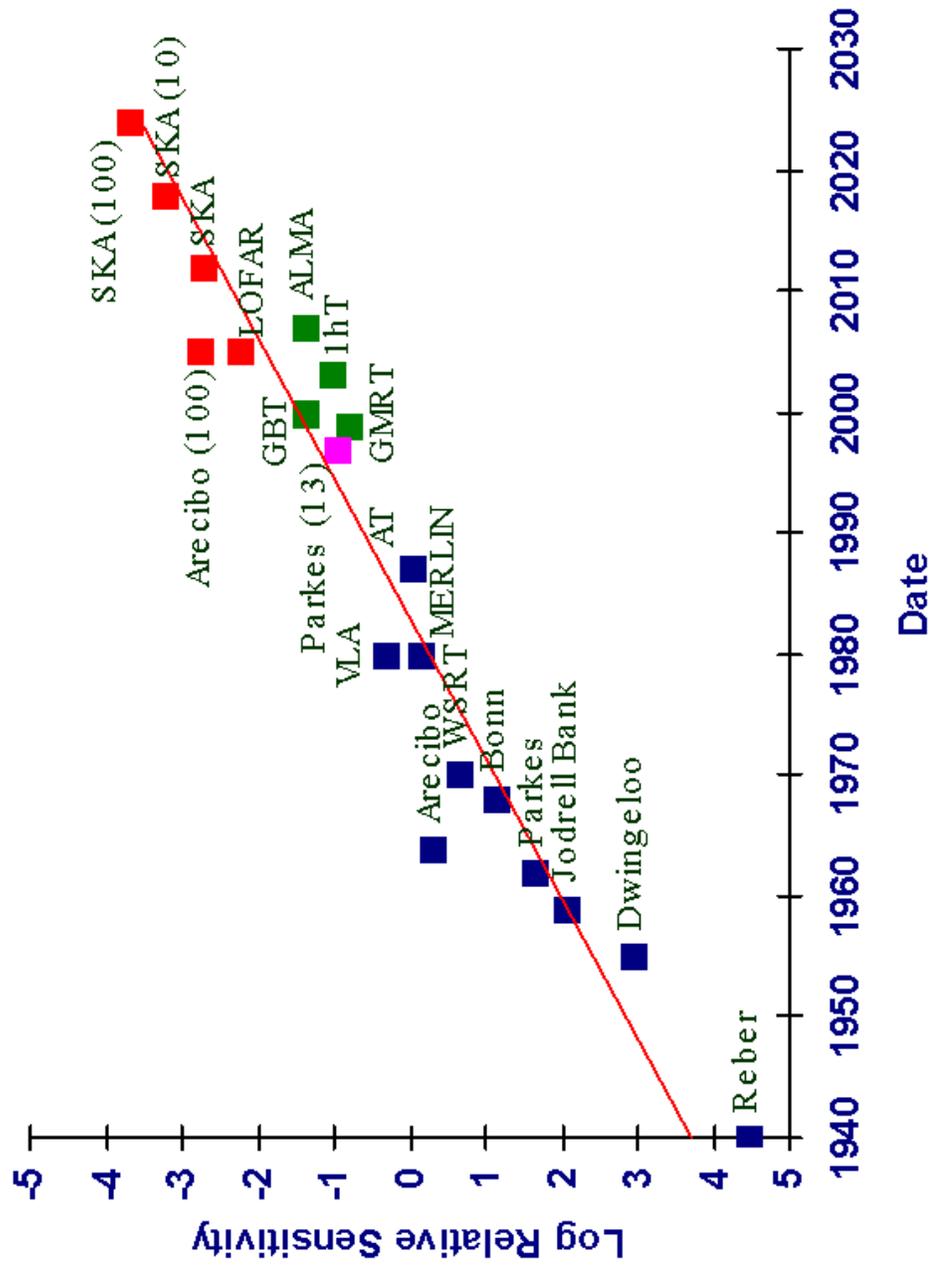}
\caption{Exponential growth of Radio Telescope sensitivity. Boxes indicate
the sensitivity attained when the systems were first commissioned. Since
then substantial improvements have been made to the system temperature and
bandwidth of many telescopes. This diagram does not convey anything about
their present capabilities. Projected capabilities are shown for ALMA,
LOFAR, SKA (with 1,10,100 beams) and Arecibo (with 100 beams). The curent 13
beams system on Parkes is also shown.} \label{fig-1}
\end{figure}

Moore's law for the growth of computing power with time (ie a doubling every
18 months) is often quoted as being vitally important for the success of the
next generation of radio telescopes (working at cm wavelengths) such as the
Square Kilometre Array (SKA). It is worth noting that radio astronomy has
enjoyed a Moore's law of it own, having a exponentially improvement in
sensitivity with time as shown in Figure \ref{fig-1}. In fact the doubling
time is approximately 3 years, and has been in progress since 1940, giving
an overall improvement in sensitivity of $10^5$. However, we are approaching
the fundamental limits of large mechanical dishes and noise limits of broad
band receivers systems. We need to look to other means of extending this
growth into the future.

Why do we want to be on this exponential growth curve ? Fields of research
continue to produce scientific advances while they maintain and exponential
growth in some fundamentally limiting parameter. For radio astronomy
sensitivity is definitely fundamentally limiting. An interesting question is
whether there are other parameters for which as exponential growth could be
maintained for a period of time. The first and most obvious point about
exponential growth is that it cannot be sustained indefinitely.  Can we
maintain it for sometime into the future ? There are 2 basic ways to stay on
the exponential curve: 1) Spend more money, 2) Take advantage of
technological advances in other areas.

\subsection{International Mega Science Projects}

International cooperation is now needed for dramatic improvements in
sensitivity, because no one country can afford to do it alone. ALMA, the
Atacama Large MM Array being developed by the USA, Europe and Japan is an
example. It will cost around \$US700M spread over 1999-2007 and will provide
an unprecedented opportunity to study redshifted molecular lines. The Square
Kilometre Array (SKA) which will work at centimetre wavelengths is likely to
be a collaboration of 10 or more nations, spending \$US500M over 2008 to
2015. 

There are 2 basic approaches to funding such large projects: 1) User pays,
where member countries pay for slice of the time, or 2) The member countries
build the facility and make it openly accessible to all. Optical astronomy
has moved very much down path 1) as the Keck, VLT, Gemini type facilities
demonstrate. Radio astronomy has traditionally followed path 2) as have
projects like CERN. For future facilities there may be pressure to move more
into the user pays regime, possibly bringing about substantial change in the
dynamics of the radio astronomy community.

In the past considerable extensibility was attained by each country
learning from the last one to build a telescope. That evolution is
very clear from Figure \ref{fig-1}. Since we are likely to be moving
to internationally funded projects, that path to extensibility is much
more restricted and designers must think very carefully about
designing extensibility into the next generation telescopes.

\subsection{Extensibility Through Improved Technologies}

For a given telescope, past and present extensibility has been
achieved by improvements in 3 main areas:

\begin{description}

\item{\bf System Temperature:} Reber started out with a 5000 K system
temperature. Modern systems now run at around 20 K, meaning that if
everything else was kept constant, Reber's telescope would now be 250
times more sensitive than when first built. There are possibilities of
some improvements in future, but nothing like what was possible in the
past.

\item{\bf Band Width:} Telescopes like the GBT (Green bank Telescope)
having bandwidths some 500 times greater than Reber's, will give
factors of 20--25 improvement in sensitivity. Some future improvements
will be possible, but again they will not be as large as in the past.

\item{\bf Multiple Beams:} Whether in the focal or aperture plane, multiple
beam systems provide an excellent extensibility path, allowing vastly deeper
surveys than were possible in the past. Although multiple beam systems have
been used for a number of experiments in the past, the full potential of
this approach is yet to be exploited. A notable example that has made a
stride forward in this direction is the Parkes L band system (Stavely-Smith
et al. 1996, PASA, 13, 243). The fully sampled focal plane phased array
system being developed at NRAO by Fisher and Bradley highlights the likely
path for the future.

\end{description}

Using these three methods, a small telescope like the Parkes 64m has
remained on the exponential curve and the forefront of scientific discovery
for 35 years (shown in Figure \ref{fig-1} in 1962 and in 1997). Other
telescopes have of course undergone similar evolution and we only high light
Parkes as an example. Scope for continuing this evolution looks good for the
next decade, but beyond that more collecting area will be needed. A 64 beam
system installed in 2010, would allow Parkes to stay on the curve for some
time. The technology to do this is probably only 3-4 years away from having
a realisable system, making it possible to jump well ahead of the
curve. Putting a 100 beam system on Arecibo by 2005 is possible and would
allow Arecibo to jump way out in front of the curve as it did when first
built in 1964.

The relevance of Moore's law in this context, is that if it continues to
hold true for the next 1-2 decades, it will provide the necessary back end
computational power to realise the gains possible with multiple beams.

\section{Key Technologies, driving future developments:}

\begin{description}
\item{\bf HEMT receivers} which are wide band, cheap, small, reliable, and
low noise systems with many elements.

\item{\bf Focal plane arrays} giving large fully sampled fields of view
will allow rapid sky coverage for survey applications and great
flexibility for targetted observations, including novel possibilities
for calibration and interference excision.

\item{\bf Interference rejection} allowing passive use of spectrum,
outside the bands allocated to passive uses. High dynamic range linear
systems, coupled with high temperature superconducting or photonic
filters will allow use of the spectrum between communication
signals. Adaptive techniques may allow some cochannel experiments, by
removing the undesired signals, so that astronomy signals can be seen
behind them.

\item{\bf More computing capacity} may result in much more of the system
being defined in software rather than hardware. This may lead to a
very different expenditure structure, where software is a capital
expense and computing harware is a considered as a consumable or
running cost.

\item{\bf Fibre/photonic} based beamforming and transmission of recorded
signals will revolutionise bandwidths and signal quality, especially
for high resolution science.

\item{\bf Software radio and smart antenna} techniques which will
allow great flexibilty in signal processing and signal selection.

\end{description}

\section{Summary of New Facilities}

\begin{table}
\begin{center}
\begin{tabular}{lllll}\hline
	& D(m)		& Area(m2)	  & Freq(GHz)	& Date \\ \hline
ALMA	& 64 x 12m	& 7.2 x $10^3$ 	  & 30.0 - 900 	& 2007 \\
GBT	& 100m	 	& 7.8 x $10^3$ 	  & 0.30 - 86	& 2000 \\ 
1hT	& 512 x 5m	& 1.0 x $10^4$ 	  & 1.00 - 12	& 2003 \\
VLA 	& 27 x 25m	& 1.3 x $10^4$ 	  & 0.20 - 50	& 2002 \\
GMRT	& 30 x 45m	& 4.8 x $10^4$	  & 0.03 - 1.5	& 1999 \\ 
SKA	& undecided	& 1.0 x $10^6$	  & 0.20 - 20	& 2015 \\
LOFAR	& $10^6$ x 1m	& 1.0 x $10^6$	  & 0.03 - 0.2	& 2003 \\ \hline
\end{tabular}
\end{center}
\end{table}

\section{Interference Sources and Spectrum Management}

It is important to be clear of what we mean when we talk about
interference. Radio astronomers make passive use of many parts of the
spectrum legally allocated to communication and other services.  As a
result, many of the unwanted signals are entirely legal and
legitimate. We will adopt the working definition that interference is
any unwanted signal, getting into the receiving system.

If future telescopes like the SKA are developed with sensitivities up
to 100 times greater than present sensitivities, it is quite likely
that current regulations will not provide the necessary
protection. There is also a range of experiments (eg redshifted
hydrogen or molecular lines) which require use of the whole spectrum,
but only from a few locations, and at particular times, suggesting
that a very flexible approach may be beneficial. Other experiments
require very large bandwidths, in order to have enough
sensitivity. presently only 1-2\% of the spectrum in the metre and
centimetre bands is reserved for passive uses, such as radio
astronomy. In the millimetre band, much larger pieces of the spectrum
are available for passive use, but the existing allocations are not
necessarily at the most useful frequencies.

\subsection{Terrestrial Sources of Interference}

Interference can arise from a wide variety of terrestrial sources,
including communications signals and services, electric fences, car
ignitions, computing equipment, domestic appliances and many others.
All of these are regulated by national authorities and the ITU
(International Telecommunications Union). In the case of Australia,
there is a single communcations authority for whole country and
therefore for the whole continent. As a result there is a single
database containing information on the frequency, strength, location,
etc of every licensed transmitter. This makes negotiations
over terrestrial spectrum use simpler in principle.

\subsection{Space \& Air Borne Sources of Interference}

Radio astronomy could deal with most terrestrial interfering signals,
by moving to a remote location, where the density and strength of
unwanted signals is greatly reduced. However with the increasing
number of space borne telecom and other communications systems in low
(and mid) Earth orbits, a new class of interference mitigation
challenges are arising - radio astronomy can run, but it cant hide !
The are several new aspects introduced to the interference mitigation
problem by this and they include: rapid motion of the transmitter,
more strong transmitters in dish sidelobes and possibly in primary
beam, and different spectrum management challenges.

There is an upside to the space borne communication systems in that
they help to develop the technology that make space VLBI possible,
which leads to the greatest possible resolution.

A classic example of the problems that can arise is provided by
Irridium mobile communications system, which has a constellation of
satellites transmitting signals to every point on the surface of the
Earth. Unfortunately in this case, there is some leakage into the
passive band around 1612 MHz, with signals levels up to $10^{11}$ times
as strong as signals from early universe.

\subsection{Radio Quiet Reserves}

Radio quiet reserves have been employed in a number of places, with
Green Bank being a notable success. For future facilties such as the
SKA and ALMA, the opportunity exists to set radio quiet reserve
planning in process a deceade before the instruments are actully
built.  Radio quiet reserves of the future may take advantage not only
of spatial and frequency orthogonality to human generated signals, but
also time, coding and other means of multiplexing. These later
parameters may be particularly important for obtaining protection from
space borne undesired signals, a number of which illuminate most of
the Earths surface.

\section{Radio Wavelength Fundamentals}

Undesired interfering signals and astronomy signals can differ (be
orthogonal) in a range of parameters:

\noindent
$\bullet$ Frequency\\
$\bullet$ Time\\
$\bullet$ Position\\ 
$\bullet$ Polarisation\\
$\bullet$ Distance\\
$\bullet$ Coding\\

It is extremely rare that interfering and astronomy signals do not possess
some level of orthogonality in this 6 dimensional parameter space. We
therefore need to develop sufficiently flexible back end systems to take
advantage of the orthogonality and separate the signals. This is of course
very similar to the kinds of problems faced by mobile communication
services, which are being addressed with smart antennas and software radio
technologies.

\section{Interference Excision Approaches}

There is no silver bullet for detecting weak astronomical signals in the
presence of undesired human generated signals. Spectral bands allocated for
passive use, provide a vital window, which cannot be achieved in any other
way. There are a range of techniques that can make some passive use of other
bands possible and in general these need to be used in combined or
complimentary way.

\begin{description}

\item{\bf Screening} to prevent signals entering the primary elements
of receivers.

\item{\bf Front end filtering} (possibly using high temperature super
conductors) to remove strong signals as soon as they enter the signal
path.

\item{\bf High dynamic range linear receivers} to allow appropriate
detection of both astronomy (signals below the noise) and interfering
signals.

\item{\bf Notch filters} (digital or ananlog) to excise particularly
bad spectral regions.

\item{\bf Decoding} to remove multiplexed signals. Blanking of period
or time dependent signals is a very succesful but simple case of this
more general approach.

\item{\bf Calibration} to provide the best possible characterisation
of interfering and astronomy signals.

\item{\bf Cancellation} of undesired signals, before correlation using
adaptive filters and after taking advantage of phase closure techniques
(Sault et al. 1997)

\item{\bf Adaptive beam forming} to steer nulls onto interfering
sources. Conceptually, this is equivalent to cancellation, but it provides a
way of taking advantage of the spatial orthogonality of astronomy and
interfering signals.

\end{description}

\subsection{Adaptive Systems}

\begin{figure}[htbp]
\plotone{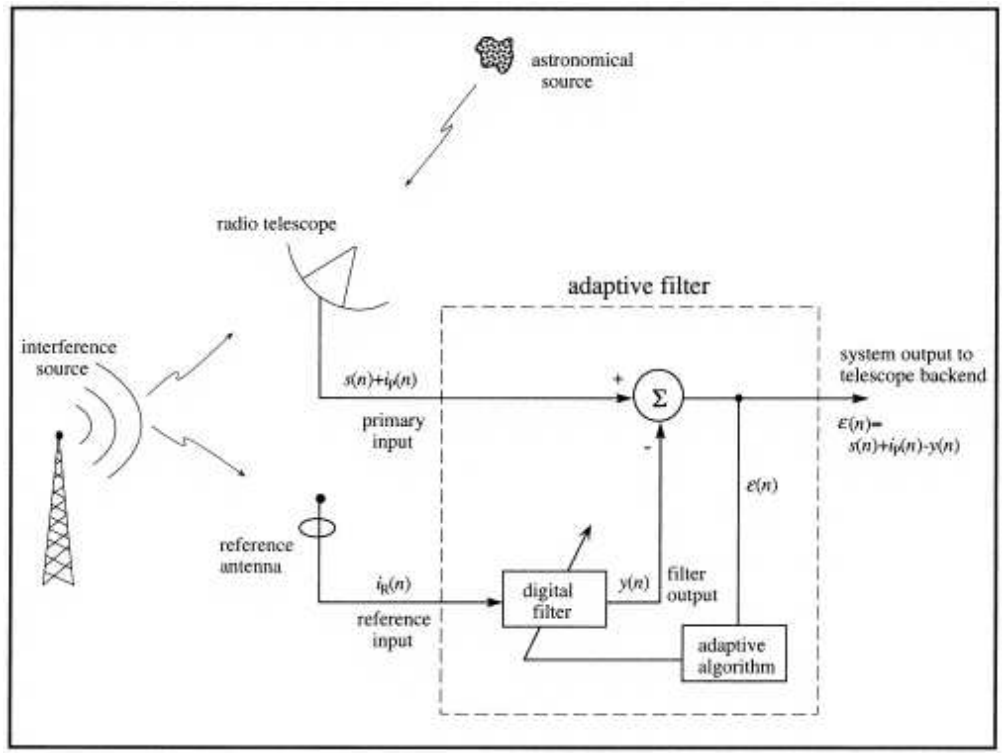}
\caption{Example of an adaptive cancellation system. From Barnbaum \&
Bradley (1998).} \label{fig-2}
\end{figure}

Of all the approaches listed above, the nulling or cancellation systems (may
be adaptive) are the most likely to permit the observation of weak astronomy
signals that are coincident in frequency with undesired signals. These
techniques have been used extensively in military, communications, sonar,
radar, medicine and others (Widrow \& Stearns 1985, Haykin 1995). Radio
astronomers have not kept pace with these developments and in this case need
to infuse rather diffuse technology in this area. A prototype cancellation
system developed at NRAO (shown in Figure \ref{fig-2}) has demonstrated 70dB
of rejection on the lab bench and 30dB of rejection on real signals when
attached to the 140 foot at Green Bank (Barnbaum \& Bradley 1998). Adaptive
Nulling systems are being prototyped by NFRA in the Netherlands. However
their application in the presence of real radio astronomy signals is yet to
be demonstrated and their toxicity to the weak astronomy signals needs to be
quantified. The best prospect for doing this in the near future is recording
baseband data from exisiting telescope, containing both interferring and
astronomy signals (Bell et al. 1999). A number of algorithms can then be
implemented is software and assessed relative to each other.

\section{The Telecommunications Revolution}

We cannot (and dont want to) impede this revolution, but we can try to
minimise its impact on passive users of the radio spectrum and maximise the
benefits of technological advances. The deregulation of this industry has
had some impact on the politics. Major companies now play a prominent
(dominate ?) role in the ITU (International Telecommunications
union). Protection of the bands for passive use must therefore addressed and
promoted by government.

\section{Spectrum Pricing}

There may be some novel ways in which spectrum pricing could evolve in
order to provide incentives for careful use of a precious
resource. Radio astronomy and other passive users cannot in general
afford commercial rates and therefore need government support. One
possibility would be to have a green tax which could be used to fund
interference management and research. 

Such strategies do not come without cost. While the long term economic
cost may be relatively small, upfront R\&D costs to an individual
company may compromise their competitiveness. This issue must
therefore be addressed at national or international policy level.

Unlike many other environmental resource use problems, spectrum over use is
both reversible and possible to curtail. This leads to certain political
advantages because politicians like to have problems which they can solve
and this is a more soluble problem than many other environemntal problems.

\section{Remedies}

\begin{description}
\item {\bf Siting Radio Telescopes} Choose remote sites with natural
shielding helps but doesn't protect against satellite
interference. Establish radio quiet zones, using National government
regulations. This is easier for fixed than mobile transmitters. Far side of
moon or L2 Lagrangian point are naturally occurring radio quite zones but
are very expensive to use.

\item {\bf OECD Mega Science Forum: Task Force on Radio Astronomy} The goals
of the OECD mega science forum are complimentary to IAU efforts, providing a
path for top down influence of governments which would otherwise not be
possible. This task force aims to promote constructive dialog between:
regulatory bodies, international radio astronomy community,
telecommunications companies, and government science agencies. It will
investigate three approaches favoured by Megascience Forum: Technological
solutions, regulation, and radio quiet reserves.

\item {\bf Environmental Impact} In the meter and cm band $<$1\% is
allocated to passive use! 99\% already used, resource use has been
extravagant. Almost all the spectrum at wavelenghts $>$ 1cm are now
polluted, and situation is rapidly deteriorating at shorter wavelengths.
It is reversable and sheareable in more creative ways! in contrast to
most other pollution problems.

\end{description}

\section{Funding of radio astronomy}

University based radio astronomy research in the USA has suffered relative
to other wavelengths for 2 reasons: 1) The centralised development at NRAO
has made it difficult for many universities to remain involved in technical
developments. 2) Space based programs in infrared, optical, UV, X-ray, and
Gamma ray bands have a rather different funding structure, where access to
research funds is based on successful observing proposals. Radio astronomy
has no access to such funds and therefore is a relatively uneconomical
pursuit for astronomers. This method of funding is being taken up in other
countries and radio astronomy needs to find a way to join the scheme.  More
globally, radio astronomy has suffered relative to other wavelengths,
because data acquisition, reduction and analysis is unnecessarily
complex. Researchers have to spend a lot more effort in data processing than
other areas of astronomy. For example, in many other bands, fully calibrated
data lands on the researchers desk a few days after the observations were
taken. Radio astronomy needs to finds ways to move into this regime (as
Westerbork have done to some degree), but at the same time preserve the vast
flexibility that can be derived from measuring the electric field at the
aperture.

\section{Conclusions}

The possibilities for the future of radio astronomy are good, but
there are some challenges issues for the community to consider and
address:

\noindent
$\bullet$ Whole of radio spectrum needed for redshifted lines\\
$\bullet$ About 2\% of spectrum is reserved for passive use by regulation
so must develop other approaches\\
$\bullet$ We cannot (and don't want to) impede the telecommunications revolution\\
$\bullet$ Radio astronomy has low credibility until we use advanced techniques\\
$\bullet$ Essential to influence government policy\\
$\bullet$ Astronomers should have a uniform position\\
$\bullet$ Threatening language doesn't help\\
$\bullet$ Is interference harmful?\\


\begin{references}

\reference Barnbaum, c. \& Bradley, R., 1998, AJ, 116, 2598.

\reference Bell J.~F., et al. 1999 "Software radio telescope: interference
      mitigation atlas and mitigation strategies", in Perspectives in Radio
      Astronomy: Scientific Imperatives at cm and m Wavelengths (Dwingeloo:
      NFRA), Edited by: M.P.  van Haarlem \& J.M. van der Hulst.

\reference Haykin, S., 1995, ``Adaptive Filter Theory'' Prentice Hall.

\reference Sault, B., Ekers, R., Kewley, L. 1997 ``Cross-correlation
approaches to interference mitigation'' Sydney SKA workshop
http://www.atnf.csiro.au/SKA/WS/

\reference Smolders, A.~B., 1999, ``Phased-array system for the next
generation of radio telescopes'', in Perspectives in Radio Astronomy:
Scientific Imperatives at cm and m Wavelengths (Dwingeloo: NFRA), Edited by:
M.P.  van Haarlem \& J.M. van der Hulst.

\reference Staveley-Smith L. et al. 1996, PASA, 13, 243 "The Parkes 21cm
Multibeam Receiver". An overview of the science and overall system design of
the multibeam receiver.

\reference Widrow, B. \& Stearns, S., 1985, ``Adaptive Signal Processing''
Prentice Hall

\reference Interference Mitigation Web pages
http://www.atnf.csiro.au/SKA/intmit/

\reference AN SKA web site http://www.atnf.csiro.au/SKA/
\end{references}
\end{document}